\documentclass[letter,scriptaddress,twocolumn, prl,showkeys]{revtex4}
  \usepackage{amsmath}
  \usepackage{amssymb}
  \usepackage{makeidx}
  \usepackage{amsfonts}
  \usepackage[ansinew]{inputenc}
  \usepackage[usenames,dvipsnames]{pstricks}
  \usepackage{epsfig}
  \usepackage{pst-grad} 
  \usepackage{pst-plot} 
  \usepackage[colorlinks,hyperindex]{hyperref}
  \usepackage{lipsum}
  \hypersetup
  {
    colorlinks,%
    citecolor=black,%
    linkcolor=black,%
    urlcolor=black,%
  }

\makeindex

\begin{document}

\title {Quantitative Study of Beam Splitter Generated Entanglement from\\
Input States with Multiple Nonclassicality Inducing Operations}

\author{Soumyakanti Bose}
\email{soumyakanti@bose.res.in}
\affiliation{S. N. Bose National Centre for Basic Sciences \\ Block-JD, Sector-III, Salt Lake, Kolkata 700106 \\ India.\\
}
\author{M. Sanjay Kumar}
\email{sanjay@bose.res.in}
\affiliation{S. N. Bose National Centre for Basic Sciences \\ Block-JD, Sector-III, Salt Lake, Kolkata 700106 \\ India.\\
}

\date{\today}

\begin{abstract}
Continuous-variable beam-splitter (BS)-generated entanglement from single-mode optical states generated by a single nonclassicality (NC)-inducing operation has been found to be immensely important in several information processing tasks. There exists a broader class of optical states, generated from successive action of multiple different NC-inducing operations, which show many intriguing nonclassical properties; however, the BS conversion of the NC for such states remains unexplored. In this work we have critically analyzed the BS-generated entanglement from such nonclassical optical states at input. Here we present a scenario where BS output entanglement becomes non-monotonic with the input NC parameters, agreeable experimentally (e.g., number of photon excitation and squeezing strength), in contrast to the previous results with states comprising a single NC-inducing operation. We explain this counter intuitive feature in terms of the competition between these two NC-inducing operations as manifest in the contours of the $Q$-functions associated with these states.
\end{abstract}


\maketitle
\section*{I. Introduction}

Quantum entanglement plays the central role in various quantum information processing tasks such as quantum communication, quantum computation, \textit{etc}. \cite{introfp1}. States having such nonlocal correlation can easily be obtained from single-mode quantum light \cite{introfp2} by using a passive linear device such as a beam-splitter (BS) \cite{introfp3}. It is known that a necessary and sufficient criterion for the output states of the BS to be entangled is that at least one of the input ports is nonclassical \cite{introfp4}. Nonclassical states are the quantum states of light for which the Glauber-Sudarshan $P$ distribution \cite{introfp5} associated with the density operator $\rho$,
\begin{equation}
\rho=\int P(\alpha,\alpha^{*})|\alpha\rangle\langle\alpha|, \label{ncdef}
\end{equation}
fails to be a true probability distribution \cite{introfp6}, where $|\alpha\rangle$ stands for a coherent state. Nonclassical states of quantized light can be generated by several nonclassicality NC-inducing operations such as photon excitation \cite{introfp7}, quadrature squeezing \cite{introfp8}, amplitude squeezing \cite{introfp9}, \textit{etc}. Thus a quantitative study of BS-generated entanglement and input NC becomes very important to obtain the desired amount of entanglement by tuning experimentally agreeable NC-inducing parameters. 

In recent times, there has been an extensive quantitative and qualitative study \cite{introfp3,introfp4,introsp1,introsp2,introsp3} of BS generated entanglement for various input nonclassical states with single NC-inducing operations such as photon excitation, quadrature squeezing, \textit{etc}. It is worth noting that a quantitative aspect of these studies \cite{introfp3,introsp1,introsp2,introsp3} is that the BS output entanglement grows monotonically with the input NC. Nonetheless, there exists a broader class of optical states \cite{introsp4,introsp5} that are generated under successive action of the two different types of NC-inducing operations, \textit{viz.}, photon excitation and quadrature squeezing. These states exhibit many intriguing nonclassical effects \cite{introsp6,introsp7,introsp8}. On the other hand, their two-mode versions are found to be more entangled than the Einstein-Podolsky-Rosen (EPR) state $\vert\psi\rangle_{\textrm{EPR}}=S_{ab}(r)\vert 0\rangle_{a}\vert 0\rangle_{b}$ \cite{introsp9} and also to improve the efficiency of teleporting coherent states \cite{introsp10}, with $S_{ab}(r)=e^{\frac{r}{2}(a^{\dagger}b^{\dagger}-ab)}$ being the two mode correlated squeezing operator. However, the BS-generated entanglement from such single-mode states remains unexplored. Consequently, it becomes imperative to analyze the BS-generated entanglement from such single-mode states of light {\em generated under multiple NC-inducing operations} in the context of quantum information processing tasks with optical resources \cite{introsp11}.

In this paper, we study the quantitative aspects of BS generated entanglement from the input single-mode photon-added squeezed vacuum state (PASVS) \cite{introsp5} and the squeezed number state (SNS) \cite{introsp4}. We find that the BS output entanglement, from input SNS, depends monotonically on the number of photon excitation and quadrature squeezing. On the other hand, for input PASVS, we observe a non-monotonic dependence on both of them. To understand the specific behavior of entanglement at the BS output, we analyze the input NC in terms of well-known measures such as the nonclassical depth, the Hilbert-Schmidt distance from the nearest coherent state and the negativity of the Wigner distribution. Our analytical and numerical results show that none of these measures characterize the nonclassical aspects of PASVS and SNS properly to account for corresponding BS output entanglement. Further, to explain this counter-intuitive result, we introduce the concept of competition between the NC-inducing operations manifest in terms of $Q$-function contours. We argue that, for PASVS, such competition leads to the non-monotonic entanglement at the BS output, while for SNS the competition is insignificant. 

In Sec. II, we briefly review the known results on BS output entanglement from input states generated by single NC-inducing operations, viz., photon addition and quadrature squeezing. In Sec. III, in contrast to the cases in Sec. II, we present the dependence of BS entanglement from input states on both the NC-inducing operations, in particular, for SNS and PASVS. In Sec. IV, we present a quantitative study of effective NC of PASVS and SNS in terms of the nonclassical depth, negativity of the Wigner distribution and the Hilbert-Schmidt distance from the nearest classical state. In Sec. V, we propose a picture of these states in terms of contours of the Husimi-Kano $Q$ function. We explain the dependence of BS output entanglement on $m$ and $r$ as arising from the competition between the two different NC-inducing operations that generate the states.

\section*{II. BS Generated Entanglement from Input States Generated under Single NC-Inducing Operations}

In this section we focus on the BS-generated entanglement from the squeezed vacuum state and the photon number state. These states can be viewed as being generated from the vacuum state under the action of a single NC-inducing operations such as (i) quadrature squeezing implemented by the squeezing operation $S(r)=e^{\frac{r}{2}(a^{\dagger 2}-a^{2})}$ and (ii) photon excitation given by $a^{\dagger m}$. In the squeezed vacuum state, $S(r)|0\rangle$, the variance in the quadrature variable $X=\frac{1}{\sqrt{2}}(a^{\dagger}+a)$, $V(X)=\langle X^{2}\rangle-\langle X\rangle^{2}$, is given by $V(X)=\frac{e^{-r}}{2}$. Hence $r$ can be taken to be a measure of squeezing in the state $S(r)|0\rangle$. It may be noted that $V(X)\leq\frac{1}{2}$ (i.e., $r\geq 0$) implies the state $S(r)|0\rangle$ is nonclassical in the sense that the Glauber-Sudarshan $P$ distribution corresponding to the state is not a true probability distribution. In view of the fact that as $r$ becomes larger and larger, $V(X)$ becomes smaller and smaller compared to $\frac{1}{2}$, it may be argued that {\em higher the value of $r$, more nonclassical the state $S(r)|0\rangle$} is.
In the same way, the photon number state $|m\rangle=\frac{a^{\dagger m}}{\sqrt{m!}}|0\rangle$ is a nonclassical state. Again, it is easily argued that the state {\em $|m\rangle$ becomes more nonclassical as $m$ increases}. This is so because, as $m$ increases, the Glauber-Sudarshan $P$ distribution becomes more and more singular compared to the $\delta$ function \cite{introfp2}. Thus, we see that at least as far as the action on the vacuum state is concerned, the two NC-inducing operations, $S(r)$ and $a^{\dagger m}$, are NC-\textit{increasing} as well, specifically in the sense that the NC of the respective state \textit{increases} as $r$ or $m$ increases. Recently, Rahimi-Keshari \textit{et al.} \cite{secII1} described the NC of any quantum process as whether it induces nonclassical effects on any classical state $|\beta\rangle$. Our notion of a NC-\textit{increasing} operation is consistent with their definition since the vacuum state can be considered a special case of $|\beta\rangle$ ($\beta=0$).

With this background, we now visit the question of how the BS output entanglement $E_{\rm{BS}}$ varies with input NC when $S(r)|0\rangle$ and $|m\rangle$ are input at one port of the BS with vacuum $|0\rangle$ at the other. A passive (lossless) 50:50 BS is represented by the transformation matrix 
\begin{equation}
\begin{pmatrix}
a_{\textrm{out}} \\
b_{\textrm{out}}
\end{pmatrix}=
\begin{pmatrix}
1/\sqrt{2} & 1/\sqrt{2}\\
-1/\sqrt{2} & 1/\sqrt{2}
\end{pmatrix}
\begin{pmatrix}
a_{\textrm{in}} \\
b_{\textrm{in}}
\end{pmatrix}
\end{equation} \label{mode_rel}

between the input and output modes. For any bi-partite pure entangled state $\vert\psi_{\rm{AB}}\rangle$, entanglement is measured by the local von Neumann entropy \cite{introfp1},
\begin{equation}
E(\vert\psi_{\rm{AB}}\rangle)=-\rm{Tr}[\rho^{\rm{r}}\ln(\rho^{\rm{r}})], \label{ent_def}
\end{equation}
where $\rho^{\rm{r}}=\rm{Tr}_{A}[\vert\psi_{\rm{AB}}\rangle\langle\psi_{\rm{AB}}\vert]$. BS-generated entanglement from input state $\vert\psi\rangle$ is thus obtained by replacing $\vert\psi_{\rm{AB}}\rangle$ by the corresponding state we get at the output of the BS.

For a single-mode number state as input, the BS output state \cite{introfp3,introsp1} becomes,
\begin{equation}
\vert m,0\rangle\xrightarrow{\rm{BS}}\frac{1}{2^{m/2}}\sum_{k=0}^{m}\begin{pmatrix}
m\\
k
\end{pmatrix}^{1/2}\vert m-k,k\rangle. \label{bs_out_fock}
\end{equation} 

Using relations (\ref{ent_def}) and (\ref{bs_out_fock}), one can obtain an analytic result for BS output entanglement for $|m\rangle$ as \cite{introfp3,introsp1}
\begin{equation}
E_{\rm{BS}}(\vert m\rangle)=-\sum_{k=0}^{m} \frac{1}{2^{m}}\begin{pmatrix}
m\\
k
\end{pmatrix}\ln{\big[\frac{1}{2^{m}}\begin{pmatrix}
m\\
k
\end{pmatrix}\big]} \label{ent_fock}
\end{equation} 

Note that in our notation $E_{\rm{BS}}(|\psi\rangle)$ is the BS output entanglement when state $|\psi\rangle$ is input at the BS.

Similarly, using the technique of reduced variance matrix for a bipartite Gaussian state \cite{rvm}, for an input single-mode squeezed vacuum state, one
has the analytic result
\begin{equation}
E_{\rm{BS}}(\vert S(r)|0\rangle)=\frac{e^{\frac{r}{2}}+1}{2}\ln [\frac{e^{\frac{r}{2}}+1}{2}]-\frac{e^{\frac{r}{2}}-1}{2}\ln [\frac{e^{\frac{r}{2}}-1}{2}],
\label{ent_svs}
\end{equation}
one can see that the $E_{\rm{BS}}$ given by Eq. (\ref{ent_svs}) is an always monotonically increasing function of $r$ since the slope
\begin{equation}
\frac{\partial}{\partial r}E_{\rm{BS}}(\vert S(r)|0\rangle)
=-\frac{e^{\frac{r}{2}}}{4}\log[\frac{1-e^{\frac{-r}{2}}}{1+e^{\frac{-r}{2}}}]
\end{equation}
is always positive for all $r>0$.

\begin{figure}[h]
\includegraphics[scale=1.5]{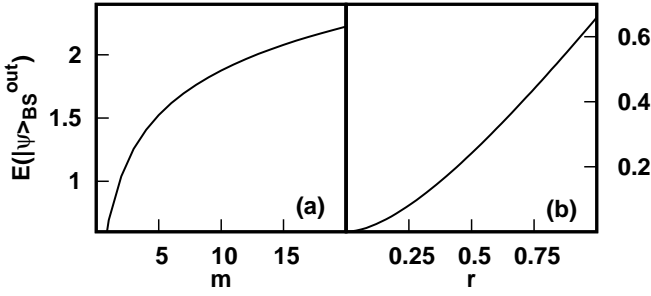}
\caption{Plot of ({\bf a}) $E_{\rm{BS}}(|m\rangle)$ with $m$ and ({\bf b}) $E_{\rm{BS}}(S(r)|0\rangle)$ with $r$}
\end{figure}

We plot the dependence of $E_{\rm{BS}}$ for input $|m\rangle$ in Fig. 1(a) and $S(r)|0\rangle$ in Fig. 1(b). Clearly, $E_{\rm{BS}}$ increases monotonically as $r$ or $m$ is increased. These results reinforce the argument made above that NC of $S(r)|0\rangle$ or $|m\rangle$ increases with $r$ or $m$ respectively because NC of the input is a resource from which one can extract entanglement in the BS setting, as has been discussed by Asboth \textit{et al.} \cite{introsp3}.

\section*{III. BS Generated Entanglement from Input States Generated under Two NC-Inducing Operations}

Let's now consider the successive action of two different NC-inducing operations, \textit{viz}., quadrature squeezing and photon addition on an initial vacuum state. This leads to SNS \cite{introsp4} and PASVS \cite{introsp5} given by
\begin{align}
\vert\psi_{\rm{SNS}}\rangle &=S(r)\vert m\rangle=\frac{S(r)a^{\dagger m}}{\sqrt{m!}}\vert 0\rangle=\sum_{k=0}^{\infty} C_{n}^{m}\vert n\rangle, \nonumber \\
\vert\psi_{\rm{PASVS}}\rangle &=\frac{1}{\sqrt{N_{m}}}a^{\dagger m}S(r)\vert 0\rangle, \label{statedef}
\end{align}
where $N_{m}=m!\mu^{m}P_{m}(\mu),P_{n}(x)$ is the $n$-order Legendre Polynomial and $C_{n}^{m}$ are given in \cite{secIII1}.

We now turn to the question of quantitative aspects of BS output entanglement when either SNS or PASVS is input at one of the ports of the BS. Using relation $(3)$ we get the BS output states for input SNS and PASVS as,
\begin{align}
\vert\psi_{\rm{SNS}}\rangle\xrightarrow{\rm{BS}}&\sum_{n=0}^{m}C_{n}^{m}\frac{1}{2^{n/2}}\sum_{p=0}^{n} \begin{pmatrix}
n\\
p
\end{pmatrix}^{1/2}\vert n-p,p\rangle, \nonumber \\
\vert\psi_{\rm{PASVS}}\rangle\xrightarrow{\rm{BS}}&\frac{1}{\sqrt{N_{m}}} \sum_{k=0}^{\infty} \frac{\sqrt{(2k+m)!}}{k!}\Big(\frac{\tau}{2}\Big)^{k}\frac{1}{2^{k+\frac{m}{2}}}\sum_{p=0}^{2k+m} \nonumber \\
&~~~~~~~~~~\begin{pmatrix}
2k+m\\
p
\end{pmatrix}^{1/2}\vert 2k+m-p,p\rangle. \label{statebsout}
\end{align}

We have plotted $E_{\rm{BS}}(\vert\psi_{\rm{PASVS}}\rangle)$ and $E_{\rm{BS}}(\vert\psi_{\rm{SNS}}\rangle)$ as a function of $r$ for various values $m$ in Fig. 2. 
\begin{figure}[h]
\includegraphics[scale=1.6]{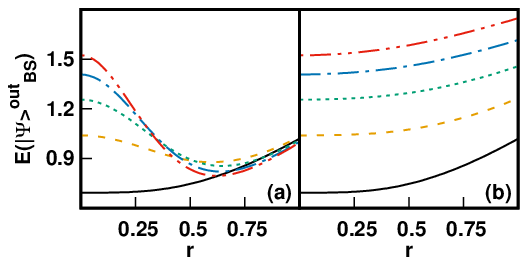}
\caption{(Color Online) Dependence of $E(\vert\Psi\rangle_{\rm{BS}}^{\rm{out}})$ for (\textbf{a}) PASVS and (\textbf{b}) SNS on $r$ for $m=1$ (black solid line), 2 (brown dashed line), 3 (green dotted line), 4 (blue dash-dotted line) and 5 (red dash-double-dotted line).}
\end{figure}

As is evident from Fig. 2(a), $E_{\rm{BS}}(\vert\psi_{\rm{PASVS}}\rangle)$ shows (except in the case of $m=1$) a {\em non-monotonic} dependence on both $r$ and $m$. For all value of $m$ ($>1$), $E_{\rm{BS}}(\vert\psi_{\rm{PASVS}}\rangle)$ first decreases and then increases with an increase in $r$. For sufficiently large $r$ in fact $E_{\rm{BS}}(\vert\psi_{\rm{PASVS}}\rangle)$ depends predominantly on $r$. Further, it can be seen from  Fig. 2(a), that because $E_{\rm{BS}}$ curves depend nonmonotonically on $r$ for various values of $m$, $E_{\rm{BS}}$ for larger values of $m$ is in fact less than that for smaller values of $m$, beyond $r \sim 0.60$.

In contrast, $E_{\rm{BS}}(\vert\psi_{\rm{SNS}}\rangle)$ [Fig. 2(b)] increases monotonically with both $r$ and $m$. This monotonic dependence is quite similar to what one has when either $S(r)|0\rangle$ or $|m\rangle$ is input at the BS as discussed in Sec. II.

From a comparison of the results in the cases of SNS and PASVS, it is evident that the dependence of BS output on $r$ and $m$ depends critically on the order in which the squeezing $S(r)$ and the photon addition $a^{\dagger m}$ operations act on the initial vacuum state. This non-monotonic dependence in the case of PASVS is indeed counterintuitive given that (as we have noted in Sec. II), BS input states generated under a single NC-inducing operation lead to monotonically increasing BS output entanglement.

It is noteworthy that in the case of $m=1$ entanglement curves for $|\psi_{\rm{PASVS}}\rangle$ and $|\psi_{\rm{SNS}}\rangle$ are identical. This feature can be understood from the following argument. For $m=1$, using the properties of the squeezing operator one can show that
\begin{align}
|\psi_{\rm{PASVS}}\rangle &=\frac{a^{\dagger}S(r)|0\rangle}{\mu}=\frac{S(r)(\mu a^{\dagger}+\nu a)|0\rangle}{\mu} \nonumber \\
&=S(r)|1\rangle=S(r)a^{\dagger}|0\rangle
\label{fid_pas_sns_1}
\end{align}
where $\mu=\cosh(r)$ and $\nu=\sinh(r)$. Thus we see that for $m=1$, $|\psi_{\rm{PASVS}}\rangle$ and $|\psi_{\rm{SNS}}\rangle$ are identical. On the other hand, for $m=2$, a similar calculation yields
\begin{align}
|\psi_{\rm{PASVS}}\rangle &=\frac{a^{\dagger 2}S(r)|0\rangle}{\mu\sqrt{2(3\mu^{2}-1)}}=\frac{S(r)(\mu a^{\dagger}+\nu a)^{2}|0\rangle}{\mu\sqrt{2(3\mu^{2}-1)}} \nonumber \\
&=\frac{S(r)(\mu^{2}a^{\dagger 2}+\mu\nu(2 a^{\dagger}a+1))|0\rangle}{\mu\sqrt{2(3\mu^{2}-1)}} \nonumber \\
&=\frac{1}{\mu\sqrt{2(3\mu^{2}-1)}}(\mu\sqrt{2}S(r)|2\rangle+\nu S(r)|0\rangle)
\label{fid_pas_sns_2}
\end{align}

Note that here $|\psi_{\rm{PASVS}}\rangle$ is a superposition of two different squeezed number states, namely, $S(r)|2\rangle$ and $S(r)|0\rangle$. For higher photon excitation ($m\geq 2$), $|\psi_{\rm{PASVS}}\rangle$ contains superposition of more SNSs and differs from the particular $S(r)|m\rangle$ even more. As a consequence, with an increase in $m$, $E_{\rm{BS}}(\vert\psi_{\rm{PASVS}}\rangle)$ differs more from $E_{\rm{BS}}(\vert\psi_{\rm{SNS}}\rangle)$ as observed in Fig. 2.

\section*{IV. Effective Single Mode NC Generated under two NC-Inducing Operations}

In a first attempt to resolve the non-monotonic dependence mentioned above, we argue that for states generated under multiple NC-inducing operations, SNS and PASVS in particular, $r$ and $m$ individually may not
measure the nonclassicality of these states, but one should perhaps work with an effective measure.  Several nonclassicality measures have been proposed in the literature such as the nonclassical depth \cite{secIV1}, Wigner negativity \cite{secIV3} and the Hilbert-Schmidt distance \cite{secIV4} from the nearest classical state. In this Sec. we shall investigate if any of these measures faithfully captures the NC of these states, and if they do, working with such effective measure (s) will allow us to understand this nonmonotonic dependence.

\subsection*{A. Nonclassical Depth}

The nonclassical depth of any quantum state of light is defined as the minimal smoothing needed to wash out the negativity (and singularity) of Glauber-Sudarshan $P$ distribution. From the $P$ distribution one can define a general $\eta$ convoluted distribution, $R(z,\eta)=\frac{1}{\eta}\int\frac{d^{2}\omega}{\pi}e^{-\frac{\vert z-\omega\vert^{2}}{\eta}}P(\omega)$. The nonclassical depth $\eta_{\rm{min}}$ is defined as the minimum value of $\eta$ needed to make $R(z,\eta)$ a positive semidefinite regular function \cite{secIV1}.

The functions $R(z,\eta)$ for PASVS and SNS (Appendix A) are given as
\begin{align}
R(z,\eta)_{\rm{PASVS}} &=\frac{A_{1}^{m}e^{\frac{\vert z\vert^{2}}{1-\eta}}W_{0}(z,z^{*},\eta)}{\mu N_{m}\sqrt{\eta^{2}-\tau^{2}(1-\eta)^{2}}}\sum_{k=0}^{m} \frac{(-1)^{m}m!}{k!(m-k)!} \nonumber \\ 
&~~~~~~~~~~~~~~~~~~~\Big(\frac{D_{1}}{A_{1}}\Big)^{k} L_{m-k}\Big(\frac{\vert B_{1}\vert^{2}}{4A_{1}}\Big), \nonumber \\
R(z,\eta)_{\rm{SNS}} &=\frac{A_{2}^{m}e^{\frac{\vert z\vert^{2}}{1-\eta}}W_{0}(z,z^{*},\eta)}{\mu\sqrt{\eta^{2}-\tau^{2}(1-\eta)^{2}}}\sum_{k=0}^{m} \frac{(-1)^{m-k}m!}{k!(m-k)!} \nonumber \\
&~~~~~~~~~~~~~~~~~~~\Big(\frac{D_{2}}{A_{2}}\Big)^{k} L_{m-k}\Big(\frac{\vert B_{2}\vert^{2}}{4A_{2}}\Big), \label{ncdepth}
\end{align}
where 
\begin{align}
&W_{0}(z,z^{*},\eta)=\exp\Big(-\frac{\frac{\eta}{1-\eta}\vert z\vert^{2}-\frac{\tau}{2}[z^{2}+z^{*2}]}{\eta^{2}-\tau^{2}(1-\eta)^{2}} \Big), \nonumber \\
&A_{1}=\frac{\tau(1-\eta)^{2}}{2[\eta^{2}-\tau^{2}(1-\eta)^{2}]}~,~A_{2}=\frac{A_{1}}{\mu^{2}}-\frac{\tau}{2}, \nonumber 
\end{align}
\begin{align}
&B_{1}=\frac{\eta z-\tau(1-\eta)z^{*}}{\eta^{2}-\tau^{2}(1-\eta)^{2}}~,~B_{2}=\frac{B_{1}}{\mu}, \nonumber \\
&D_{1}=\frac{\eta(1-\eta)}{\eta^{2}-\tau^{2}(1-\eta)^{2}}~,~D_{2}=\frac{D_{1}}{\mu^{2}}, \label{depthconst}
\end{align}
$\mu=\cosh{r},~\tau=\tanh{r},~$ and $L_{n}(x)$ is the $n\rm{th}$ order Laguerre polynomial. 

Because of the presence of the Laguerre polynomial [Eqn. (\ref{ncdepth})], the positiveness of the function $R(z,\eta)$ is not guaranteed for all choices of $\eta$. In such cases, as prescribed in \cite{secIV1}, the nonclassical depth has to be taken to be unity. Thus we have a situation where the nonclassical depth for both $|\psi_{\textrm{SNS}}\rangle$ and $|\psi_{\textrm{PASVS}}\rangle$ is the same as the photon number state and hence it is independent of the squeezing strength $r$. Clearly, the nonclassical depth fails to be a faithful measure of NC as far as these states are concerned. Further, our conclusion, specifically in the context of these states, is in agreement with the general conclusion that nonclassical depth is always unity for all non-Gaussian pure states \cite{secIV2}.

\subsection*{B. Negativity of Wigner Function}

It is well known that the Wigner function $W_{\rho}(z,z^{*})$ of any state of light $\rho$, being negative in phase space, indicates that the state $\rho$ is nonclassical. This criterion of course fails for Gaussian states. The phase space integral of the negative part of the Wigner function, the Wigner negativity
\begin{equation}
\delta=\frac{\int\frac{d^{2}z}{\pi}\vert W_{\rho}(z,z^{*})\vert-1}{2},
\end{equation}
may be considered a measure of NC \cite{secIV3}. A larger $\delta$ implies that the state is more nonclassical.

The Wigner functions of PASVS and SNS (Appendix B) are given by,
\begin{align}
W_{\rm{SNS}}(\alpha,\alpha^{*})=&(-1)^{m}e^{-2\vert \beta\vert^{2}}L_{m}(4\vert \beta\vert^{2}), \nonumber \\
W_{\rm{PASVS}}(\alpha,\alpha^{*})=&\frac{2(-1)^{m}m! e^{-2\vert \beta\vert^{2}}\mu^{m}\nu^{m}}{2^{m}N_{m}}\sum_{k=0}^{m}\frac{m!(\frac{\tau}{2})^{-k}}{k!(m-k)!} \nonumber \\
&~~~~~~~~~~~~~~~~~~~~~~L_{m-k}\Big(\frac{2|\beta|^{2}}{\tau}\Big), \label{wigneg} 
\end{align}
where $\mu=\cosh{r},\nu=\sinh{r},\beta=\mu \alpha-\nu \alpha^{*}$. 
\begin{figure}[h]
\includegraphics[scale=1.8]{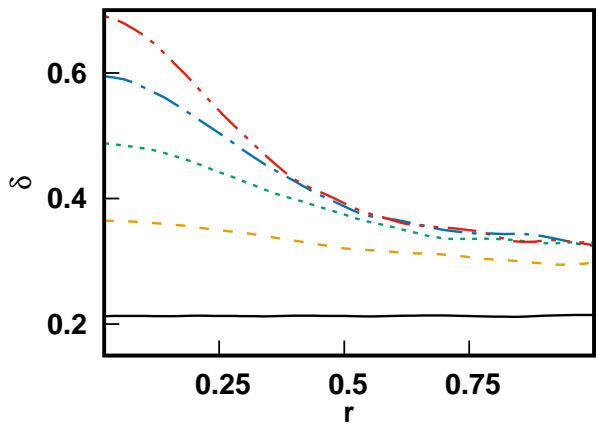}
\caption{(Color Online) Dependence of $\delta$ on $r$ for $m$=1 (black solid line), 2 (brown dashed line), 3 (green dotted line), 4 (blue dash-dotted line) and 5 (red dash-double-dotted line) for PASVS.}
\end{figure}

In Fig. 3, the Wigner negativity of PASVS is plotted as a function of the squeezing strength $r$ for various values of the photon addition number $m$. For all values of $m$, the Wigner negativity falls off with increasing
$r$. This can be understood as being due to the photon addition operation for large $r$. In the case of SNS, however, Wigner negativity is independent of $r$ and hence it is the same as that of the number state $|m\rangle$. The independence of Wigner negativity on $r$, in the case of SNS, can be easily understood from the following fact. If $W_{\rho}(x,p)$ is the Wigner function of a given state $\rho$, then the Wigner function of the state
$\rho'=S(\zeta)\rho S^{\dagger}(\zeta)$ is $W_{\rho}(x',p')$, where $x',p'$ and $x,p$ are related to each other by a linear canonical transformation. Since the Jacobian of any linear canonical transformation is unity, the
Wigner negativity of $\rho\rightarrow\rho'$ remains the same.

As in the case of the nonclassical depth, again, we have a situation where the Wigner function negativity fails to be a faithful measure of NC as far as SNS is concerned. As our aim is to do a comparative study of PASVS and SNS in the context of entanglement of the BS output state with these states as the input, it is desirable that we have a measure of NC that works equally well for both the NC-inducing operations, i.e., photon excitation and quadrature squeezing.

\subsection*{C. Hilbert-Schmidt Distance From Nearest Coherent State}

A measure of NC based on the Hilbert-Schmidt distance between density operators has been proposed in the literature \cite{secIV4}. This measure is defined as the Hilbert-Schmidt distance of a given density operator $\rho$ from the nearest classical state. Since coherent states $|\beta\rangle$ are the only pure classical states $|\beta\rangle$ \cite{secIV5}, $d_{\rm{NC}}$ for a pure state $|\psi\rangle$ is defined as,
\begin{equation}
d_{\rm{NC}}=\inf\sqrt{2}[1-|\langle\beta\vert\psi\rangle|^{2}]^{1/2}.
\end{equation}
where infimum is taken over the set of all coherent states $|\beta\rangle$, with $\beta$ being a complex number.

We have calculated $d_{\rm{NC}}$ for the two states PASVS and SNS. While $d_{\rm{NC}}$ for PASVS has a closed-form analytic expression given by 
\begin{align}
d_{\rm{NC}}\Big)_{\rm{PASVS}}=&\sqrt{2}\Big[1-\frac{m^{m}e^{-m}}{(1-\tau)^{m}N_{m}}\Big]^{\frac{1}{2}}, \label{ncdistpasvs}
\end{align}
$d_{\rm{NC}}$ for SNS can at best be reduced to the simple form
\begin{align}
d_{\rm{NC}}\Big)_{\textrm{SNS}}=&\inf_{\beta}\sqrt{2}\Big[1-\frac{\tau^{m}e^{-\vert\beta\vert^{2}+\frac{\tau}{2}(\beta^{2}+\beta^{*2})}}{\mu 2^{m}} L_{m}\Big(\frac{\vert\beta\vert^{2}}{2\mu\nu}\Big)\Big]^{\frac{1}{2}}, \label{ncdistsns}
\end{align}
from which the computation proceeds via a numerical optimization. 

In Figs. 4(a) and 4(b) we have plotted $d_{\rm{NC}}$ for PASVS and SNS. For PASVS, $d_{\rm{NC}}$ first decreases and then increases with an increase in $r$ for all $m$; however, for $r\lesssim 0.20$, we observe a monotonic dependence of $d_{\rm{NC}}$ upon $m$ while for larger $r (\gtrsim 0.20)$ such monotonicity breaks down. For $m\geq 2$, $d_{\rm{NC}}$ shows a non-monotonic behavior [Fig. 4(a)] consistent with that of $E_{\rm{BS}}$ [Fig. 2(a)]. In contrast, in the case of $m=1$, $d_{\rm{NC}}$ reveals non-monotonic behavior that is inconsistent with $E_{\rm{BS}}$. Here, we have a situation, in particular, for $m=1$, in the case of PASVS, where {\bf NC} (as measured by $d_{\rm{NC}}$) decreases while the corresponding $E_{\rm{BS}}$ [Fig. 2(a)] increases, which is unphysical. On the other hand, for SNS, we observe a non-monotonic dependence of $d_{\rm{NC}}$ [Fig. 2(b)] on $r$ but a monotonic dependence on $m$. For all $m$, as $r$ increases, it first decreases and then increases. Similar to the case for PASVS, for SNS, we also have an unphysical situation where the NC (as measured by $d_{\rm{NC}}$) decreases while $E_{\rm{BS}}$ [Fig. 2(b)] increases with $r$ for all $m$. This leads us to the conclusion that $d_{\rm{NC}}$ is not an acceptable measure of NC of states generated under multiple NC-inducing operations, in particular SNS and PASVS.
\begin{figure}[h]
\includegraphics[scale=1.6]{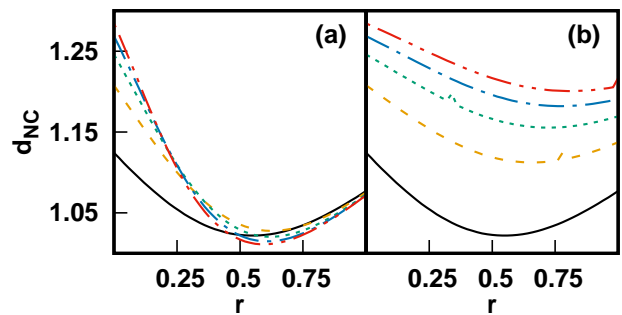}
\caption{(Color Online) Dependence of $d_{\rm{NC}}$ for (\textbf{a}) PASVS and (\textbf{b}) SNS on $r$ for $m$=1 (black solid line), 2 (brown dashed line), 3 (green dotted line), 4 (blue dash-dotted line) and 5 (red dash-double-dotted line)}. 
\end{figure}

It appears from the above discussion that none of the three measures considered above is an acceptable measure of NC of the states we have studied in this paper. Whether a suitable measure of {\bf NC} can be given that shows a dependence on $r$ and $m$ for such states that is consistent with the dependence of $E_{\rm{BS}}$ on these parameters remains {\em an open question}. 

\section*{V. Monotonicity Versus NonMonotonicity question; Role of Competing Nonclassicalities}

In this Sec. we shall outline our point of view that the nonmonotonicity in the $E_{BS}$ curves (in the case of PASVS) is a consequence of a {\em competition} between the two different kinds of NC-inducing operations underlying these states. Various counterintuitive features seen in Fig. 2(a) can be attributed to the effects of such a competition, in particular the feature that we discussed after Fig. 2, i.e.., that $E_{\rm{BS}}$ for larger values of $m$ is in fact less than that for smaller values of $m$ beyond $r \sim 0.60$.

We illustrate this competition in terms of contours of the $Q$ function associated with these states.  To begin with, it is helpful to visualize the effect of the two NC-inducing operations acting individually on an initial vacuum state in terms of the deformation induced in the circular $Q$ function contour of the vacuum state.  As is well known, light (initially in a coherent state) propagating through a medium with a Kerr nonlinearity undergoes radial squeezing \cite{rsq} and the photon number state can be thought of as an extreme case of a radially squeezed state. Here, figuratively speaking, the photon excitation (addition) operation $(a^{\dagger})^{m}$ can be thought of as deforming the circular $Q$ function contour of the vacuum state into an extreme crescent shape. On the other hand, the squeezing operation $S(r)$ can be thought of as deforming the initial circular $Q$ function contour of the vacuum state into an ellipse \cite{esq}.

The above picture can now be applied to states with two NC-inducing operations applied in succession. As is evident from Fig. 5, in the case of PASVS, for small $r$, with an increase in $m$, the contours become more crescent shaped indicating the dominant number state character. However, as $r$ increases, except for the case of $m=1$, the crescent-shaped contours smooth out and become more elliptic. This points to a crossover in the dominant character of the state, from a photonic to a quadrature squeezed one. Such a crossover arises due to an {\em overwhelming competition} between photon addition and quadrature squeezing operations. For higher $r$ $(\gtrsim 0.60)$, the $Q$ function contours tend to become more and more elliptic. It is our view that this competition, as manifest in terms of the crossover from crescent-shaped to elliptic contours of the $Q$ function, is what is behind the change in slope that is evident in the $E_{\rm{BS}}$ curve for PASVS [Fig. 2(a)].
\begin{figure}[h]
\includegraphics[scale=1.4]{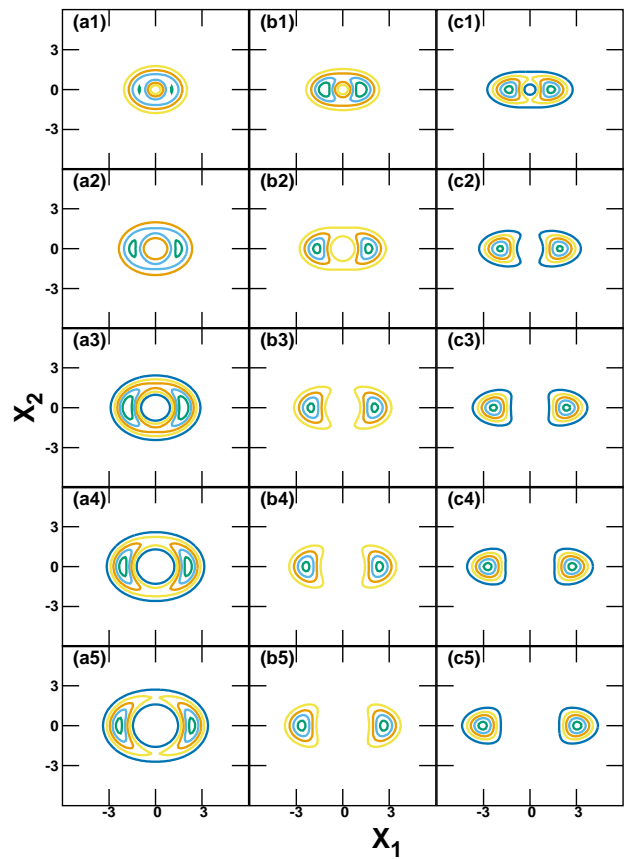}
\caption{(Color Online) Contour plots of the $Q$ function for PASVS for different $m$ and $r$. The axes of the subplots are the quadrature components given by $X_{1}=\frac{\beta+\beta^{*}}{\sqrt{2}}$ and $X_{2}=\frac{\beta-\beta^{*}}{i\sqrt{2}}$.}
\end{figure} 

On the other hand, in the case of SNS (Fig. 6), unlike in the case of PASVS, there does not appear to be any significant crossover from crescent-shaped to predominantly elliptic $Q$-function contours. This points to a rather {\em insignificant competition} between the two NC-inducing operations, namely, photon excitation and quadrature squeezing.  Consequently, no change in slope is evidenced in the corresponding $E_{\rm{BS}}$ curve [Fig. 2(b)].

To sum up, the key to understanding the monotonicity vs nonmonotonicity question is therefore the degree of competition between the two NC inducing operations. An overwhelming competition leads to a slope change in the $E_{\rm{BS}}$ curve and hence a non-monotonic dependence. Whether this competition is overwhelming or insignificant can be inferred from the contours of the $Q$ functions associated with the states depending on whether or not they undergo a crossover from crescent shaped to elliptic, as $r$ or $m$ is increased.
\begin{figure}[h]
\includegraphics[scale=1.4]{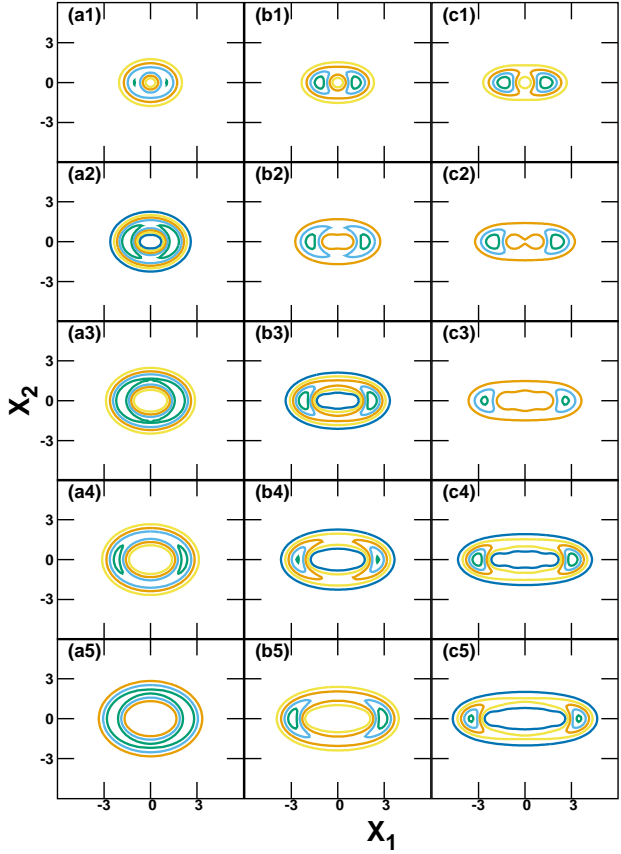}
\caption{(Color Online) Contour plots of the $Q$ function for SNS for different $m$ and $r$. The axes are the same as in Fig. 5.}
\end{figure}

\section*{VI. Conclusion}

In conclusion, we have quantitatively studied the BS output entanglement for states generated from successive application of two different NC-inducing operations that lead to, in particular, SNS and PASVS. We have observed that while BS output entanglement shows a monotonic dependence on the squeezing strength and the number of photon addition in the case of SNS, this dependence is nonmonotonic in the case of PASVS. We show that any attempt to understand this issue of monotonicity vs nonmonotonicity fails since none of the measures such as the nonclassical depth, the Wigner negativity and the Hilbert-Schmidt distance proves to be an acceptable measure of NC of these states. We have offered an intuitive picture in terms of contours of the associated $Q$ function of these states and pointed out that the competition between these two different NC-inducing operations is the key to understand the monotonicity vs nonmonotonicity issue.

\section*{Appendix A: $R(z,\eta)$ for PASVS and SNS}

Introducing the expression of $P(\gamma)$ in terms of density operator $\rho$, the $\eta$ convoluted function $R(z,\eta)$ can be written in terms of $\rho$ as,
\begin{equation}
R(z,\eta)=\frac{e^{\frac{|z|^{2}}{1-\eta}}}{1-\eta}\int\frac{d^{2}\beta}{\pi}\langle -\beta|\rho|\beta\rangle e^{-\frac{(2\eta-1)\vert \beta\vert^{2}+(z^{*}\beta-z\beta^{*})}{1-\eta}}.
\label{R_def}
\end{equation}

For PASVS, we have $\langle\beta|\psi_{\textrm{PASVS}}\rangle=\frac{\beta^{*m}}{\sqrt{\mu N_{m}}}e^{-\frac{|\beta|^{2}}{2}+\frac{\tau}{2}\beta^{2}}$. Thus, the $R(z,\eta)$ for PASVS becomes,
\begin{align}
R(z,\eta)_{\rm{PASVS}}&=\frac{1}{1-\eta}e^{\frac{\vert z\vert^{2}}{1-\eta}}\int \frac{d^{2}\beta}{\pi}\langle -\beta\vert\psi_{\rm{PASVS}}\rangle\langle\psi_{\rm{PASVS}}\vert \beta\rangle \nonumber \\
&~~~~\exp\Big(-\frac{(2\eta-1)\vert \beta\vert^{2}+(z^{*}\beta-z\beta^{*})}{1-\eta}\Big) \nonumber \\
&=\frac{1}{\mu N_{m}(1-\eta)}e^{\frac{\vert z\vert^{2}}{1-\eta}}\int \frac{d^{2}\beta}{\pi}(-\vert \beta\vert^{2m}) \nonumber \\
&\exp\Big(-\frac{\eta |\beta|^{2}}{1-\eta}+\frac{\tau (\beta^{2}+\beta^{*2})}{2}+\frac{z\beta^{*}-z^{*}\beta}{1-\eta}\Big) \nonumber 
\end{align}

One can derive the above non-Gaussian integral using parametric differentiation as,
\begin{widetext}
\begin{align}
R(z,\eta)_{\rm{PASVS}}&=\frac{(-1)^{m}}{\mu N_{m}(1-\eta)}e^{\frac{\vert z\vert^{2}}{1-\eta}}\partial_{a}^{m}\partial_{b}^{m}\Big[\exp\Big(-\frac{\eta}{1-\eta}\vert \beta\vert^{2}+\frac{\tau}{2}(\beta^{2}+\beta^{*2})-\frac{z^{*}}{1-\eta}\beta+\frac{z}{1-\eta}\beta^{*}\Big) \nonumber \\
&~~~~~~~~~~~~~~~~~~~~~~~~~~~~~~~~~~~~~~~~~~~~~~~~~~~~~~~~~~~~~~~~~~~~~~~~~~~~~\exp\Big(a\beta+b\beta^{*}\Big)\Big]_{a=0,b=0} \nonumber \\
&=\frac{(-1)^{m}}{\mu N_{m}(1-\eta)}e^{\frac{\vert z\vert^{2}}{1-\eta}}\partial_{a}^{m}\partial_{b}^{m}\Big[\int \frac{d^{2}\beta}{\pi} \exp\Big(-\frac{\eta}{1-\eta}\vert \beta\vert^{2}+\frac{\tau}{2}(\beta^{2}+\beta^{*2})\Big)  \nonumber \\
&~~~~~~~~~~~~~~~~~~~~~~~~~~~~~~~~~~~~~~~~~~~~~~~~~~~\exp\Big((a-\frac{z^{*}}{1-\eta})\beta+(b+\frac{z}{1-\eta})\beta^{*}\Big)\Big]_{a=0,b=0} \nonumber 
\end{align}
\end{widetext}

For any Gaussian integral, we know that
\begin{equation}
\int\frac{d^{2}z}{\pi}e^{\zeta |z|^{2}+\xi z+\eta z^{*}+f z^{2}+g z^{*2}}=\frac{e^{\frac{-\zeta\xi\eta+f\eta^{2}+g\xi^{2}}{\zeta^{2}-4fg}}}{\sqrt{\zeta^{2}-4fg}}.
\label{gauss_int}
\end{equation}

provided $\zeta^{2}-4fg >0$. Using formula (\ref{gauss_int}), for $R(z,\eta)_{\rm{PASVS}}$, we get,
\begin{widetext}
\begin{align}
R(z,\eta)_{\textrm{PASVS}}& =\frac{A_{1}^{m}e^{\frac{\vert z\vert^{2}}{1-\eta}}}{\mu N_{m}\sqrt{\eta^{2}-\tau^{2}(1-\eta)^{2}}}W_{0}(z,z^{*},\eta)\partial_{a}^{m}\partial_{b}^{m}\Big[ e^{A_{1} a^{2}+B_{1} a-B_{1}^{*}+D_{1} a b+A_{1} b^{2}}\Big]_{a=0,b=0} \nonumber \\
& =\frac{A_{1}^{m}e^{\frac{\vert z\vert^{2}}{1-\eta}}}{\mu N_{m}\sqrt{\eta^{2}-\tau^{2}(1-\eta)^{2}}}W_{0}(z,z^{*},\eta) \sum_{k=0}^{m} (-1)^{m} \frac{m!}{k!(m-k)!}\Big(\frac{D_{1}}{A_{1}}\Big)^{k} L_{m-k}\Big(\frac{|B_{1}|^{2}}{4A_{1}}\Big). 
\label{R_pasvs}
\end{align}
where,
\begin{eqnarray}
W_{0}(z,z^{*},\eta)=\exp\Big(-\frac{\frac{\eta}{1-\eta}\vert z\vert^{2}-\frac{\tau}{2}(z^{2}+z^{*2})}{\eta^{2}-\tau^{2}(1-\eta)^{2}} \Big), ~A_{1}=\frac{\tau(1-\eta)^{2}}{2[\eta^{2}-\tau^{2}(1-\eta)^{2}]} \nonumber \\
B_{1}=\frac{\eta z-\tau(1-\eta)z^{*}}{\eta^{2}-\tau^{2}(1-\eta)^{2}},~ D_{1}=\frac{\eta(1-\eta)}{\eta^{2}-\tau^{2}(1-\eta)^{2}}
\end{eqnarray}
\end{widetext}

Similarly, using the technique of parametric differentiation, one can easily derive $R(z,\eta)$ for an SNS. Since $\langle\beta|\psi_{\rm{SNS}}\rangle=\frac{e^{-\frac{|\beta|^{2}}{2}+\frac{\tau}{2}\beta^{*2}}}{\sqrt{\mu m!}}\partial_{a}^{m}\Big[ e^{-\frac{\tau}{2}a^{2}+\frac{\beta^{*}}{\mu}a}\Big]_{a=0}$, we have

\begin{widetext}
\begin{align}
R(z,\eta)_{\rm{SNS}} &=\frac{1}{1-\eta}e^{\frac{\vert z\vert^{2}}{1-\eta}}\int \frac{d^{2}\beta}{\pi}\langle -\beta\vert\psi_{\rm{SNS}}\rangle\langle\psi_{\rm{SNS}}\vert \beta\rangle \exp\Big(-\frac{(2\eta-1)\vert \beta\vert^{2}+(z^{*}\beta-z\beta^{*})}{1-\eta}\Big) \nonumber \\
& =\frac{e^{\frac{\vert z\vert^{2}}{1-\eta}}}{\mu m! (1-\eta)}\partial_{a}^{m}\partial_{b}^{m} \Big\lbrace \exp\Big(-\frac{\tau}{2}(a^{2}+b^{2})\Big) \nonumber \\
&~~~~~~~~~~~~~~~~~~~~~\int \frac{d^{2}\beta}{\pi} \exp\Big[-\frac{\eta}{1-\eta}\vert \beta\vert^{2}+(\frac{b}{\mu}-\frac{z^{*}}{1-\eta})\beta-(\frac{a}{\mu}-\frac{z}{1-\eta})\beta^{*}+\frac{\tau}{2}(\beta^{2}+\beta^{*2})\Big]\Big\rbrace_{a=0,b=0} \nonumber \\
& =\frac{A_{2}^{m}e^{\frac{\vert z\vert^{2}}{1-\eta}}}{\mu\sqrt{\eta^{2}-\tau^{2}(1-\eta)^{2}}}W_{0}(z,z^{*},\eta)\partial_{a}^{m}\partial_{b}^{m}\Big[ e^{A_{2} a^{2}+B_{2} a+B_{2}^{*}-D_{2} a b+A_{2} b^{2}}\Big]_{a=0,b=0} \nonumber \\
& =\frac{A_{2}^{m}e^{\frac{\vert z\vert^{2}}{1-\eta}}}{\mu\sqrt{\eta^{2}-\tau^{2}(1-\eta)^{2}}}W_{0}(z,z^{*},\eta)\sum_{k=0}^{m} (-1)^{m-k}\frac{m!}{k!(m-k)!}\Big(\frac{D_{2}}{A_{2}}\Big)^{k} L_{m-k}\Big(\frac{|B_{2}|^{2}}{4A_{2}}\Big).
\label{R_sns}
\end{align}
where, \begin{equation}
W_{0}(z,z^{*},\eta)=exp\Big(-\frac{\frac{\eta}{1-\eta}\vert z\vert^{2}-\frac{\tau}{2}(z^{2}+z^{*2})}{\eta^{2}-\tau^{2}(1-\eta)^{2}} \Big),~A_{2}=\frac{A_{1}}{\mu^{2}}-\frac{\tau}{2},~B_{2}=\frac{B_{1}}{\mu},~D_{2}=\frac{D_{1}}{\mu^{2}}
\end{equation}
\end{widetext}

\section*{Appendix B: $W(z,z^{*})$ for PASVS and SNS}

Here, using the technique discussed in appendix A, we calculate the Wigner function for PASVS and SNS. The Wigner distribution for any density operator is given as,
\begin{equation}
W(z,z^{*})=2 e^{2|z|^{2}}\int \frac{d^{2}\beta}{\pi}\langle -\beta\vert\psi_{\textrm{SNS}}\rangle\langle\psi_{\textrm{SNS}}\vert \beta\rangle e^{2(z\beta^{*}-z^{*}\beta)}.
\label{wig_def}
\end{equation}

Thus, the Wigner distributions for PASVS and SNS are given as
\begin{align}
W_{\rm{PASVS}}(\alpha,\alpha^{*})&=2e^{2|\alpha|^{2}}\int\frac{d^{2}\beta}{\pi}\langle -\beta|\psi_{\rm{PASVS}}\rangle \nonumber \\
&\langle\psi_{\rm{PASVS}}|\beta\rangle \exp\Big(2(\alpha\beta^{*}-\alpha^{*}\beta)\Big) \nonumber
\end{align}
\begin{widetext}
\begin{align}
&=\frac{2e^{2|\alpha|^{2}}}{N_{m}\mu}\int\frac{d^{2}\beta}{\pi}e^{-|\beta|^{2}+2(\alpha\beta^{*}-\alpha^{*}\beta)}\partial_{p}^{m}\Big[\int\frac{d^{2}\gamma}{\pi}e^{-|\gamma|^{2}-\beta^{*}\gamma+p\gamma^{*}+\frac{\tau}{2}\gamma^{*2}}\Big]_{p=0} \partial_{q}^{m}\Big[\int\frac{d^{2}\eta}{\pi}e^{-|\eta|^{2}+q\eta+\beta\eta^{*}+\frac{\tau}{2}\eta}\Big]_{q=0} \nonumber \\
&=\frac{2e^{2|\alpha|^{2}}}{N_{m}\mu}\partial_{p}^{m}\partial_{q}^{m}\Big[\frac{1}{\sqrt{1-\tau^{2}}}e^{\frac{1}{1-\tau^{2}}\Big(-(p-2\alpha)(q-2\alpha^{*})+\frac{\tau}{2}\overline{(p-2\alpha)^{2}+(q-2\alpha^{*})^{2}} \Big)}\Big]_{p=0,q=0} \nonumber \\
&=\frac{2e^{2[(\mu^{2}+\nu^{2})|\alpha|^{2}-\mu\nu(\alpha^{2}+\alpha^{*2})]}}{N_{m}}\partial_{p}^{m}\partial_{q}^{m}\Big[e^{\frac{\mu\nu}{2}(p^{2}+q^{2})-\mu^{2}pq+2\mu\overline{q(\mu\alpha-\nu\alpha^{*})+p(\mu\alpha^{*}-\nu\alpha)}} \Big]_{p=0,q=0} \nonumber \\
&=\frac{2(-1)^{m}m! e^{-2\vert \beta\vert^{2}}\mu^{m}\nu^{m}}{2^{m}N_{m}}\Sigma_{k=0}^{m}\frac{m!(\frac{\tau}{2})^{-k}}{k!(m-k)!} L_{m-k}\Big(\frac{2|\beta|^{2}}{\tau}\Big). \label{wig_pasvs}
\end{align}
\begin{align}
W_{\textrm{SNS}}(\alpha,\alpha^{*})&=2e^{2|\alpha|^{2}}\int\frac{d^{2}\beta}{\pi}\langle -\beta|\psi_{\textrm{SNS}}\rangle\langle\psi_{\textrm{SNS}}|\beta\rangle e^{2(\alpha\beta^{*}-\alpha^{*}\beta)} \nonumber \\
&=\frac{2e^{2|\alpha|^{2}}}{m!\mu}\partial_{p}^{m}\partial_{q}^{m}\Big[ e^{-\frac{\tau}{2}(p^{2}+q^{2})}\int\frac{d^{2}\beta}{\pi}e^{-|\beta|^{2}+(\frac{q}{\mu}-2\alpha^{*})\beta-(\frac{p}{\mu}-2\alpha)\beta^{*}+\frac{\tau}{2}(\beta^{2}+\beta^{*2})} \Big]_{p=0,q=0} \nonumber \\
&=\frac{2e^{2[(\mu^{2}+\nu^{2})|\alpha|^{2}-\mu\nu(\alpha^{2}+\alpha^{*2})]}}{m!}\partial_{p}^{m}\partial_{q}^{m}\Big[ e^{-pq+2q(\mu\alpha-\nu\alpha^{*})+2p(\mu\alpha^{*}-\nu\alpha)} \Big]_{p=0,q=0} \nonumber \\
&=2(-1)^{m}e^{-2\vert \beta\vert^{2}}L_{m}(4\vert \beta\vert^{2}), \label{wig_sns}
\end{align}
\end{widetext}
where $\mu=\cosh r,\nu=\sinh r,\beta=\mu \alpha-\nu \alpha^{*}$. Evidently, for $\eta=\frac{1}{2}$, Eqs. (\ref{R_pasvs}) and (\ref{R_sns}) coincide with Eqs. (\ref{wig_pasvs}) and (\ref{wig_sns}) respectively, since for $\eta=\frac{1}{2}$, $R(z,\eta)$ coincides with the Wigner function, $W(z,z^{*})$.

\end{document}